\documentclass[twocolumn]{article}
\usepackage{graphicx,amsmath}

  \renewenvironment{thebibliography}[1]{%
    \begin{oldthebibliography}{#1}%
      \setlength{\parskip}{0ex}%
      \setlength{\itemsep}{0ex}%
  }%
  {%
    \end{oldthebibliography}%

  }

\begin{document}

\title{A Unified Model of Codon Reassignment in Alternative Genetic Codes}

\author{Supratim Sengupta and Paul G. Higgs$\footnote{Corresponding Author. E-mail : higgsp@mcmaster.ca}$\\
Department of Physics and Astronomy \\ 
Mcmaster University, Hamilton, Ontario, Canada L8S 4M1\\
\\ 
}

\maketitle

{\bf \small Many modified genetic codes are found in specific genomes in which one or more codons have been 
reassigned to a different amino acid from that in the canonical code. We present a new framework for codon 
reassignment that incorporates two previously proposed mechanisms (codon disappearance and ambiguous intermediate) 
and introduces two further mechanisms (unassigned codon and compensatory change). Our theory is based on the 
observation that reassignment involves a gain and a loss. The loss could be the deletion or loss of function of 
a tRNA or release factor. The gain could be the gain of a new type of tRNA, or the gain of function of an 
existing tRNA due to mutation or base modification. The four mechanisms are distinguished by whether the 
codon disappears from the genome during the reassignment, and by the order of the gain and loss events. We 
present simulations of the gain-loss model showing that all four mechanisms can occur within the same 
framework as the parameters are varied. We investigate the way the frequencies of the mechanisms are influenced 
by selection strengths, the number of codons undergoing reassignment, directional mutation pressure, and 
selection for reduced genome size. }
\vspace{5 pt}

 The genetic code is a mapping between the 64 possible codons in mRNAs and the 20 amino acids. The canonical genetic code 
was established before the last universal common ancestor (LUCA) of archaea, bacteria and eukaryotes. It was initially believed 
to be universal, however many deviations from the canonical code have now been discovered in specific groups of organisms 
and organelles \cite{knight}. Changes involve the reassignment of a codon or a group of associated codons from one amino acid to another, 
from a stop codon to an amino acid, or vice versa. Reassignments are caused by changes in tRNAs or Release Factors (RF). In 
many cases, the molecular details are known, but the mechanism by which the changes were fixed in the population is more 
puzzling. If the translation apparatus changes so that a codon is reassigned to a new amino acid, this will introduce an amino 
acid change in every protein where this codon occurs. We would expect this to be a selective disadvantage to the organism, and 
hence selection should act to prevent the change in the code spreading through the population. The situation seems even worse 
for stop codons, where a change would lead either to premature termination of translation or to read-through. In spite of these 
perceived problems, a majority of the observed reassignments involve stop codons.

  Osawa and Jukes \cite{osawa} proposed that the deleterious effect of codon reassignment could be avoided if the codon first disappeared 
from the genome. Changes in the translational system would occur when the codon was absent, and when the codon reappeared it 
would be translated as a different amino acid. Following \cite{yarus1}, we will call this the codon disappearance (CD) mechanism. 
Schultz and Yarus \cite{yarus1} proposed an alternative ambiguous intermediate (AI) mechanism that does not require the codon to 
disappear. Instead, the codon is translated ambiguously as two different amino acids during the period of reassignment. Some 
cases of ambiguous translation are known \cite{matsugi,santos}, and it is possible that other cases have passed through such 
an ambiguous intermediate stage.
	
\subsection*{The Gain-Loss Framework}

 In this paper, we present a unified model for codon reassignment that includes both CD and AI mechanisms as special cases, and 
introduce two other possible mechanisms. We refer to this as the gain-loss model, as it is based on our observation that codon 
reassignments involve a gain and a loss. By loss, we mean the deletion of the gene for the tRNA or RF originally associated with 
the codon to be reassigned, or the loss of function of this gene due to a mutation or base modification in the anticodon. By gain, 
we mean the gain of a new type of tRNA for the reassigned codon (e.g. by gene duplication), or the gain of function of an existing 
tRNA due to a mutation or a base modification enabling it to pair with the reassigned codon.

	We will consider two specific examples. Firstly, in the canonical code, AUU, AUC and AUA are all Ile, and AUG is Met. 
In animal mitochondria, AUA is reassigned to Met. Most organisms require two tRNAs to translate Ile codons. One tRNA has a GAU 
anticodon, and this translates AUU and AUC codons. The other has a K2CAU anticodon, where K2C is Lysidine, a modified C base 
that is capable of pairing with A in the third position of the AUA codon. The Met tRNA has CAU anticodon and pairs only with AUG. 
The Lysidine modification is known in E. coli \cite{muramatsu}, and this tRNA gene is also found in all available complete genomes of 
proteobacteria \cite{tang} and in several protist mitochondrial genomes \cite{gray}. The same mechanism is presumed to operate in 
all these cases. The Lysidine tRNA is absent in animal mitochondrial genomes, and the tRNA-Met is modified so that it can pair 
with both AUG and AUA codons. This can happen either by a mutation to UAU or by base modification to f$^5$CAU \cite{tomita}. Thus, 
in our model, the loss represents the deletion of the K2CAU gene and the gain represents the gain of function of the tRNA-Met. The 
second example is the reassignment of UGA from stop to Trp. This involves the loss of the RF binding to this codon \cite{tomita}, 
and the gain of function of the tRNA-Trp by mutation and subsequent base modification from a CCA anticodon to U*CA, where U* is a 
modified U which pairs with purines but not with pyrimidines.

\begin{figure}[th!]
\vspace*{-2.2cm}
\includegraphics[width=3.2in,angle=0]{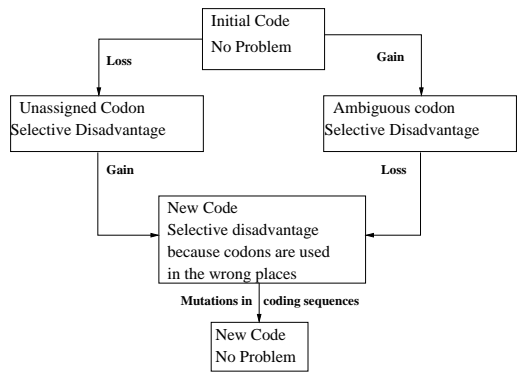}
\vspace*{-3.0cm}
\caption{A schematic picture of the codon reassignment process interpreted in terms of
the gain-loss framework. The strength of selective disadvantage depends on the number of
times a codon is used. There is no selective disadvantage if the codon disappears.}
\label{schema}
\end{figure}

Although the details of the molecular event causing the gain and loss differ for each case of codon reassignment, the
common gain-loss framework (see Fig.\ref{schema})applies to all of them. We begin with an organism following the 
canonical code and end with an organism with a modified code, in which both gain and loss have occurred. For the CD 
mechanism, the change is initiated by the disappearance of the codon undergoing reassignment prior to the fixation of loss 
or gain. The codon remains absent during the intermediate stage of reassignment until the fixation of a subsequent gain or 
loss event leads to the establishment of the new code. In that case, the gain and loss events will be selectively neutral, 
because in absence of the codon, the changes in the tRNAs make no difference to the organism. Therefore, the temporal 
order of gain and loss events is irrelevant. If the codon does not disappear the order of gain and loss events is relevant. 
In the AI mechanism, the gain occurs before the loss. Two tRNAs can associate with the codon, and hence the codon will 
sometimes be mistranslated, causing a selective disadvantage for the organism. If the loss then occurs, the codon is no 
longer ambiguous, and a new code has been established. However, there is still a selective disadvantage because the organism 
is using the codon in places where the old amino acid is required. As mutations to other synonymous codons occur, this 
disadvantage will gradually disappear. The reassigned codon can also begin to appear in positions where the new amino acid 
is required, and this will not cause a selective disadvantage.

  A third possible mechanism is characterized by the loss occurring first, creating a state where there is no specific 
tRNA for the codon. The subsequent occurrence of a gain establishes the new code. We call this the unassigned codon (UC) 
mechanism. In the unassigned state there will be a selective disadvantage because translation will be inefficient. We presume 
that, after loss of the tRNA, some other tRNA has some degree of affinity for the codon, and that eventually an amino acid 
will be added for this codon permitting normal translation to continue. In some cases \cite{yokobori}, loss of a specific tRNA can 
occur with little penalty because another more general tRNA is available that can pair quite well with the codon in question. 
For example in Echinoderm, Hemichordate and Platyhelminthes mitochondria, an Ile tRNA with GAU anticodon can pair with the 
AUA codon in absence of the specific tRNA (with Lysidine in the wobble position) mentioned above. Also, in canonical code, 
AGA and AGG are Arg, but the corresponding tRNA-Arg has been lost from animal mitochondrial genomes. In some phyla, the AGR codons 
are unassiged and are absent from the genomes. In other phyla, a gain has also occurred: the Ser tRNA is modified to either UCU 
or m$^7$GCU, which pairs with all four AGN codons \cite{matsuyama}. In the Urochordates, there has been a gain of a new tRNA-Gly 
specific to AGR codons \cite{kondow}. Thus several different responses to the loss of the original tRNA-Arg have occurred.	

  The gain-loss framework described here is related to the model of compensatory mutations \cite{kimura}. This model considers two genes 
with two alleles 0 and 1, such that both genes are in the 0 state initially. Each of the 1 alleles has a deleterious effect 
when present alone. When both mutations occur together, they compensate for one another and the fitness of the 11 state is 
assumed to be equal to the 00 state. Kimura showed that pairs of compensatory mutations can accumulate relatively rapidly, and 
this is one form of neutral evolution. A specific place where compensatory mutations are frequent is in the paired regions of 
RNA secondary structures. A population genetics theory for compensatory substitutions in RNA has been given by 
Higgs \cite{higgs}. A key point is that the pair of mutations can either be fixed at two successive times or 
simultaneously. For example, if the population is initially GC, a mutation at one site can create a GU pair 
that goes to fixation as a slightly deleterious allele, and a second mutation can create an AU pair that 
compensates for the first change and then becomes fixed. Alternately, if there is a significant selective 
disadvantage to the GU pair, it is likely to remain at very low frequencies in the population and not go 
to fixation. If a second mutation occurs in a GU sequence, this creates an AU sequence that can spread 
as a compensatory neutral pair. In this case both the A and the U mutations become fixed simultaneously 
and there is no period where the GU variant spreads throughout in the population. When real RNA sequences 
are analyzed \cite{savill}, the rate of double substitutions is high compared to single substitutions, 
suggesting that this mechanism of simultaneous fixation of compensatory pairs is frequent. This brings 
us to the fourth possibility that can occur within our framework, which we call the compensatory change 
(CC) mechanism. We presume that gain and loss events are occurring continually in genomes at some low 
rate, creating individuals with a selective disadvantage because of ambiguous or unassigned codons. If 
this disadvantage is substantial, these variants are unlikely to reach high frequencies in the population. 
However, if the compensatory gain or loss occurs in one of these individuals, the gain and loss may then 
spread simultaneously through the population.

  This model concerns historical changes in the code that occurred after the establishment of the canonical code. It is 
likely that there were also prehistoric changes that led to the establishment of the canonical code before the LUCA. 
During this prehistoric stage, the code may have had fewer than 20 amino acids, and changes may have involved the
addition of new amino acids to the code. The coevolution theory \cite{wong} considers the way the increase in 
amino acid diversity in the code was correlated to the synthesis of new amino acids. It has been argued 
\cite{szathmary} that codon swapping reassignments driven by directional mutation pressure could have played a 
role in establishing the canonical code. Moreover, the canonical code appears to be optimized so as to reduce the 
deleterious effects of non-synonymous substitutions. Tests have shown that the canonical code is better than 
almost all randomly reshuffled codes in this respect \cite{freeland1,freeland2}. A specific model for code build up and 
optimization has been investigated \cite{ardell} with the aim of understanding the pattern of amino acid 
associations of codons and the observed redundancy of the canonical code.It is likely that reassignments of 
sense codons or capture of non-sense codons could occur more easily during the early period of evolution when 
various pathways of establishing an optimal code were being explored. On the other hand, the negative selective 
effects of a reassignment process after the optimal canonical code has been established is likely to be very high. 
This puts severe restrictions on the possible reassignments that can occur in the post-canonical code era.

 Although optimization of the code may have been important in the establishment of the canonical code, we do not 
believe that historical codon reassignments have been driven by selection for changes that improve on the 
canonical code. {\it All} the present day alternative codes are much better than random codes, and the variant 
codes seem to be slightly worse than the canonical code \cite{freeland2}.  Our view is that reassignments arise 
due to chance gain and loss events that become fixed in the population. It is the strength of the negative 
selective effects acting during the period of reassignment that controls the rate of reassignment and the 
mechanism by which it occurs. Possible differences in fitness between the codes before and after completion 
of the reassignment are most likely negligible with respect to the selective effects occurring during 
the reassignment.

   Having described the four possible mechanisms within our framework, we now present a population genetics model in 
which all four of these mechanisms can be shown to occur. We use simulations to investigate the relative frequency 
of the mechanisms.

\subsection*{The Model}

  We consider an asexual, haploid population of $N$ individuals, each possessing a genome of $L$ DNA bases, representing 
protein-coding regions. Non-coding regions are ignored. A target protein sequence is specified consisting of $L/3$ amino 
acids or stops. The DNA sequences of the population evolve under mutation, selection and drift. The fitness of a DNA sequence 
is determined by how close its specified protein is to the target. Translation of the DNA is performed initially according 
to the canonical code, but when the code changes different organisms in the population may be translated by different codes. 
For every amino acid that is different from the target amino acid (mis-sense mutations) there is a factor $(1-s)$ in the fitness. 
There is a factor $(1-s_{stop})$ in the fitness for every stop where an amino acid should be and for every amino acid where a 
stop should be (nonsense mutations). The fitness of a sequence with j mis-sense and k nonsense mutations is 
$w = (1-s)^j(1-s_{stop})^k $. 

 Each individual has a parent chosen randomly from the previous generation with a probability proportional to its fitness. 
The DNA sequence of the offspring is copied from the parent. Mutations occur with a probability $u$ per site. Each base is 
replaced by one of the other three bases with equal probability. Each individual begins with a genome that exactly specifies 
the target protein. As mutations accumulate, the population reaches a state of mutation-selection balance where an individual 
typically differs from the target at a few points. (Note that we choose parameter values where the population is not subject to 
Muller's ratchet).

\begin{figure}[t!]
\vspace{-0.5in}
\includegraphics[width=3.3in,angle=0]{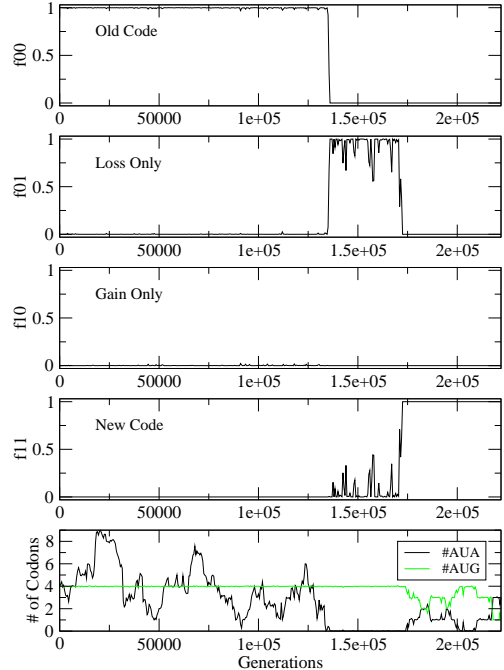}
\vspace{-0.5in}
\caption{Evidence of reassignment of AUA via the CD mechanism.
Variation of frequeny of the four types of individuals are shown as a function of time. 
The bottom panel shows the mean number of AUA \& AUG codons per genome.
The parameters are $s=s_{stop}=s_{unas}=0.05$, $u=0.0001$, and $\#AUA=4$ in the
original sequence.}
\label{cdaua}
\end{figure}

  In each run we consider only one codon reassignment, and we specify in advance which codon will be reassigned. Each individual 
possesses a Loss gene and a Gain gene with two alleles, 0 or 1. If $Loss = 0$, the original tRNA is still present and functional 
in the genome. If $Loss = 1$, the original tRNA has been deleted or has become dysfunctional due to a deleterious mutation. If 
$Gain = 0$, a new tRNA is not yet available to translate the codon. If $Gain = 1$, a tRNA capable of translating the codon with a 
new meaning has been acquired. There are four types of individuals defined by the Gain and Loss genes. Type $00$ individuals 
use the old genetic code. Type $11$ individuals use the new code and the DNA sequence is translated according to the new code 
at the point when its fitness is calculated. Type $01$ individuals have an unassigned codon. A penalty of $(1 - s_{unas})$ is 
associated with each unassigned codon in a genome. Type $10$ individuals have an ambiguous codon. Roughly half of the occurrences 
of this codon will be wrongly translated. Therefore a fitness contribution of $(1-s)^{1/2}$ is associated with such a codon whenever 
it appears in a position where the target is either the old or the new amino acid. If the codon appears in a position where the 
target is neither of these, it will always be wrong; therefore, a factor of $(1-s)$ applies.

  The probabilities of a gain and a loss occurring per generation per whole population are $U_g$ and $U_l$. If a gain or loss 
occurs then it is implemented in one random individual. Mutation occurs from the 0 to the 1 allele in each case and we do not 
consider reversals of these changes. Thus, the simulation will eventually end up with all individuals in the $11$ state. However, 
for some parameter values the population remains in the $00$ state for as long as it is feasible to run the simulations. The results 
presented below are for parameter values where the codon reassignment occurred in a reasonable time. By following the sequence of 
gain and loss events in our in silico experiment, we are able to determine the mechanism responsible for the reassignment of a codon 
and the frequency of occurrence of each mechanism. When counting the number of times each mechanism occurs, the mechanism is defined 
as CD if the codon disappears from the population during the reassignment period, irrespective of the order of the gain and loss. 
The underlying mechanism is CC if the gain and loss are fixed simultaneously. If the codon does not disappear, the mechanism is 
AI if the gain is fixed first and UC if the loss is fixed first.

\begin{figure}[t!]
\vspace{-0.5in}
\includegraphics[width=3.3in,angle=0]{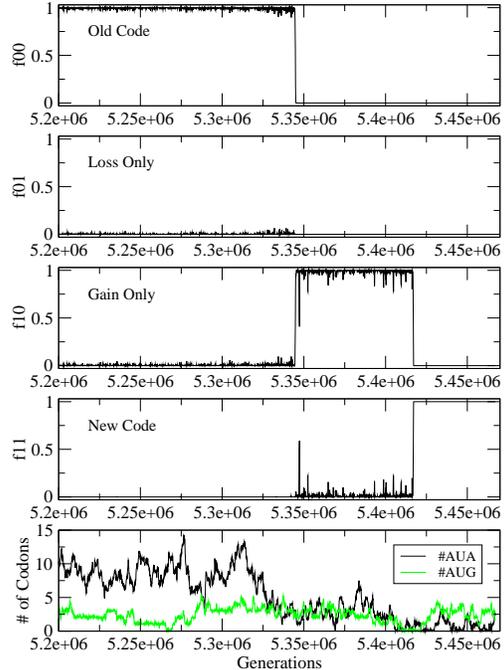}
\vspace{-0.5in}
\caption{Evidence of reassignment of the AUA codon via the AI
mechanism. The parameters $s=s_{stop}=s_{unas}=0.01$, u=0.0001, and \#AUA=12 in
the original sequence.}
\label{aiaua}
\end{figure}

\subsection*{Results}

   The following examples concern the reassignment of the AUA codon. The number of Ile codons in the target sequence was set 
to 12, 24 or 36. There were 4 codons for each of the 19 other amino acids plus a single stop codon. Thus the total number of 
codons was 89, 101, or 113. The genome lengths used were quite short due to constraints on computing time. The simulations 
were initiated with equal numbers of AUA, AUU and AUC codons (i.e. \#AUA = 4, 8 or 12 initially). The initial DNA sequence for the 
non-Ile codons was chosen randomly from codons corresponding to the target. The population size was $N = 1000$, and $U_g = U_l = 0.1$. 
The other parameters were varied between simulations.

	Fig. \ref{cdaua} shows an example of the CD mechanism. The frequency of Type $ij$ individuals is denoted $f_{ij}$. The 
simulation begins with $f_{00}=1$ and ends with $f_{11}=1$. There is an intermediate stage where $f_{01}$ goes to unity, 
indicating that the loss happens first. However, in the bottom panel it is clear that \#AUA goes to zero just before $f_{01}$ 
increases. Hence, the reassignment process is initiated by the disappearance of the AUA codons. In this example, the selection 
against mis-sense mutations, $s$, is much greater than the mutation rate, $u$. This means that the mean number of codons for 
each amino acid is almost exactly equal to the number in the target sequence. As AUG is the only Met codon, \#AUG stays fixed 
close to 4 until the codon reassignment occurs. In contrast, \#AUA can fluctuate due to third position mutations. 
After the change, both AUA and AUG code for Met, consequently \#AUA + \#AUG = 4.

	Fig. \ref{aiaua} shows a reassignment occurring by the AI mechanism. \#AUA is increased to 12 initially, which makes it less 
likely for the codons to disappear due to random drift. Moreover, $s$ is reduced, which makes the penalty for the AI state less 
severe. The gain happens before the loss. \#AUA does not disappear during the intermediate stage, however it remains around 4 
during most of the AI stage - substantially less than the 12 AUA codons originally present. This suggests that selection is acting 
against the ambiguous codon during the intermediate stage. Another effect of reduction of $s$ is that the number of Met codons fluctuates 
about its optimum value of 4, in contrast to its behaviour in Fig. \ref{cdaua}.
 
\begin{figure}[t!]
\vspace{-0.5in}
\includegraphics[width=3.2in,angle=0]{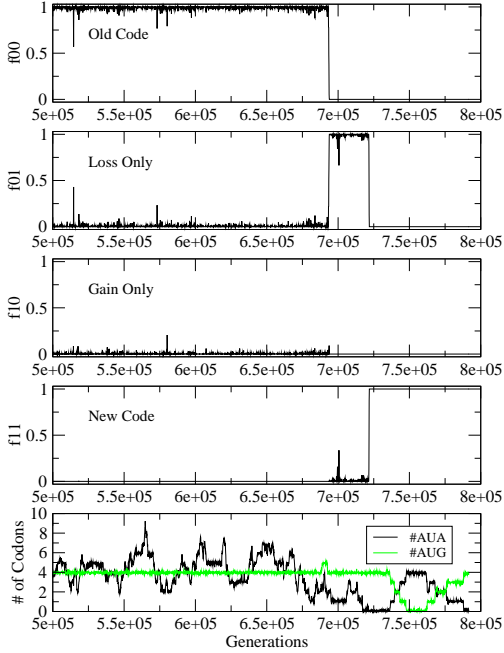}
\vspace{-0.5in}
\caption{Evidence of reassignment of the AUA codon via the UC
mechanism. The parameters $s=s_{stop}=0.02$, $s_{unas}=0.007$, 
u=0.0001, and \#AUA=4 in the original sequence.}
\label{ucaua}
\end{figure}

\begin{figure}[t!]
\vspace{-0.5in}
\includegraphics[width=3.2in,angle=0]{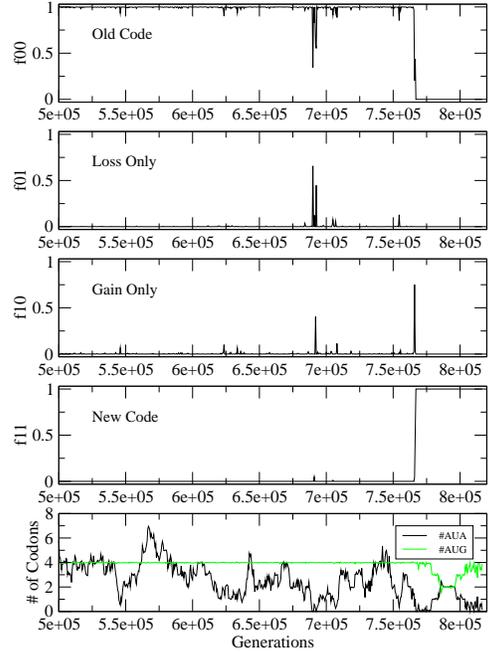}
\vspace{-0.5in}
\caption{Evidence of reassignment of the AUA codon via the compensatory change 
mechanism. The parameters $s=s_{stop}=s_{unas}=0.05$, u=0.0001, and \#AUA=4 in
the original sequence.}
\label{ccaua}
\end{figure}

 Fig. \ref{ucaua} shows the reassignment of AUA via the UC mechanism initiated by the loss of the tRNA-Met with Lysidine in the 
wobble position. The penalty against the unassigned codon is smaller than the penalty against a mis-sense or nonsense translation. 
This would be the case if another tRNA were available to translate the unassigned codon reasonably well in the absence of the original 
specific tRNA. The loss occurs whilst the \#AUA is non-zero, leaving the AUA codons unassigned. \#AUA is seen to fluctuate during 
the intermediate stage (when $f_{01}\sim1$) and eventually goes to zero during the later part of the intermediate stage. This is a 
manifestation of selection acting to minimize the use of the unassigned codon after the disappearance of its specific tRNA. However, 
since the number of codons is small, the disappearance could also be due to random drift. For lower values of $s_{unas}$, 
the \#AUA codons remains non-zero during the entire intermediate stage. The frequencies of the four types of individuals are 
shown as a function of generations in the top four panels. The bottom panel shows the mean number of AUA and AUG codons per genome.  
 Fig. \ref{ccaua} provides evidence of the reassignment of AUA from Ile to Met via the compensatory change mechanism. There is no 
intermediate period and neither $f_{01}$ nor $f_{10}$ becomes close to one. The new code replaces the old code when the gain 
and loss are fixed in the population simultaneously.

  We carried out many runs of the simulation in order to investigate the frequency of the different reassignment mechanisms. For 
each set of parameters, 20 runs were carried out using different random number seeds. Table 1(a) shows the effect of varying $s$. 
These examples correspond to AUA reassignment, and in this case the stop codon is not particularly relevant. Therefore we will set 
$s_{stop} = s$, to reduce the number of parameters. To determine the value of $s_{unas}$, we must consider two factors. There will 
be a penalty for the delay involved in waiting for a poorly matching tRNA to come along, and there will be a penalty due to incorrect 
translation of the codon in cases where the wrong amino acid is inserted. For this set of runs, we will set $s_{unas} = s$, for 
simplicity. This is appropriate if an incorrect amino acid is inserted for every occurrence of the unassigned codon, but the delay 
effect is negligible. When $s=0.05$, almost all changes occur via the CD mechanism. As selection is reduced, there is a shift 
from predominantly CD to predominantly AI. This makes sense, because with weaker selection there is a smaller penalty against 
ambiguous codons. The UC mechanism is less frequent than the AI mechanism, because of an asymmetry in the way we assigned fitness 
to ambiguous and unassigned codons. We supposed that the ambiguous codon was translated incorrectly only half the time, so that the 
penalty was $(1-s)^{1/2}$ per ambiguous codon, whereas the penalty for an unassigned codon was $(1-s)$. Additionally, we see that 
the CC mechanism occurs occasionally, and that the frequency of CC appears to be higher for larger $s$ (although the number of 
runs is too small to demonstrate this statistically). We would expect the CC mechanism to be most relevant under conditions where 
it is difficult for either the gain or the loss to be fixed in the population individually, i.e. when there is a strong penalty 
against both the ambiguous and unassigned states. The mean fixation time for the establishment of the new code was also found to 
increase with increasing $s$, as expected, since the deleterious effects of strong negative selection make it more difficult to 
implement code change.

\begin{table}[t]
\caption{Results refer to AUA codon reassignment, except for (c) which is for UGA. 
(a) Varying selection strength $s$. \#AUA=4,$u=0.0001$,$s=s_{stop}=s_{unas}$ in all cases.
(b) Varying the number of Ile codons. $u=0.0001$, $s=s_{stop}=s_{unas}=0.01$.
(c) Varying $s_{stop}$ with $s=s_{unas}=0.01$, \#UGA=4 in the original sequence.
(d) Varying GC mutational pressure. \#AUA=8, $u=0.0001,s=s_{stop}=s_{unas}=0.01$.
(e) Varying $s_{unas}$ with $u=0.0001, s=s_{stop}=0.02$, \#AUA=4.
(f) Varying selective advantage for genome size reduction by tRNA deletion ($\delta$).
$u=0.0001, s=s_{stop}=0.02$, $s_{unas}=0.007$, \#AUA=4.}
\vskip 2mm
\begin{tabular}{|c|c|c|c|c|c|} \cline{1-6}
 & & CD & UC & AI & CC \\ \cline{1-6}
(a) & $s=0.05$ & 18 & 0 & 0 & 2 \\
    & $s=0.02$ & 13 & 0 & 6 & 1 \\
    & $s=0.01$ & 4 & 3 & 13 & 0 \\ \cline{1-6}
(b) & \#AUA=4 & 4 & 3 & 13 & 0 \\
    & \#AUA=8 & 2 & 2 & 16 & 0 \\
    & \#AUA=12 & 1 & 2 & 17 & 0 \\ \cline{1-6}
(c) & $s_{stop}=0.01$ & 0 & 3 & 16 & 1 \\
    & $s_{stop}=0.02$ & 13 & 3 & 3 & 1 \\
    & $s_{stop}=0.04$ & 13 & 7 & 0 & 0 \\ \cline{1-6}
(d) & $\pi_G=0.25$ & 2 & 2 & 16 & 0 \\
    & $\pi_G=0.30 $ & 4 & 2 & 14 & 0 \\
    & $\pi_G=0.40 $ & 12 & 1 & 5 & 2 \\ \cline{1-6}
(e) & $s_{unas}=0.02$ & 12 & 0 & 5 & 3 \\
    & $s_{unas}=0.007$ & 6 & 11 & 3 & 0 \\
    & $s_{unas}=0.003$ & 1 & 19 & 0 & 0 \\ \cline{1-6}
(f) & $\delta=0$ & 6 & 11 & 3 & 0 \\
    & $\delta=0.001$ & 7 & 8 & 2 & 3 \\
    & $\delta=0.004$ & 1 & 18 & 1 & 0 \\ \cline{1-6}
\end{tabular}
\end{table}

   Table 1(b), shows the effect of varying the number of Ile codons. As described above, \#AUA initially was 4, 8 or 12. As 
\#AUA increases, there is a trend away from the CD mechanism toward the AI mechanism. The mean fixation time for the 
new code via the AI mechanism increases with \#AUA, since the cumulative deleterious effects of negative selection on many 
codons make it more difficult to establish the new code. The time taken for the CD mechanism also increases with the number 
of codons undergoing reassignment, because it is more difficult for a larger number of such codons to disappear by chance. 
However, the time for the CD mechanism increases faster, hence the shift from CD to AI with increasing codon number.

   As reassignment of UGA from stop to Trp is one of the most frequently observed reassignments, we wished to consider this 
case specifically in the model. The target sequence used in this case had 93 codons for a full range of amino acids, plus 4 
stop codons. Each of the four stop codons was set to UGA in the initial DNA sequences. The loss gene represents a RF, which 
recognizes UGA as a stop codon, and the gain gene represents the Trp tRNA. We begin in Table 1(c) with $s_{stop} = 0.01$, i.e. 
the same as $s$, and consider the effect of increasing $s_{stop}$ while $s$ is fixed. The choice of $s_{unas}$ depends on what 
we would expect to occur if the stop codon became unassigned. In absence of a specific RF, would translation terminate anyway 
because another less specific release factor was available, or would read through occur? In the latter case the selection against 
an unassigned stop codon should be as high as the selection against a nonsense translation ($s_{stop}$), but in the former case, 
it should be less. For this simulation, we set $s_{unas} = s$, for simplicity, and in absence of clear biological knowledge of 
what happens in this case. If the gain precedes the loss, an ambiguous intermediate stage is encountered during which UGA is 
read simultaneously as both stop and Trp. In this case, a penalty of $(1-s_{stop})^{1/2}$ is associated with the UGA codon 
whenever it corresponds to a stop or a Trp in the target protein sequence. As $s_{stop}$ increases, the mechanism 
changes from predominantly AI to predominantly CD. Since the penalty against an unassigned stop codon was kept fixed 
($s_{unas} =0.01$), a significant number of changes occur via the UC mechanism for large values of $s_{stop}$. However, if the 
penalty against an unassigned UGA had been as large as $s_{stop}$, CD would be the only viable mechanism. We emphasize that even in large 
genomes containing many genes, the number of UGA codons can be quite small if UAR are the primary codons used for translation 
termination. Under such circumstances, the reassignment of UGA can happen relatively easily via the CD mechanism and this provides 
a plausible reason for the large number of observed reassignments involving the UGA codon.  

   For the runs in part (c), the CD mechanism takes place when \#UGA goes to zero due to random drift. However, directional mutation 
pressure can also cause the disappearance of a codon \cite{osawa}. In the case of the AUA reassignment, biased mutation away from A  
can cause AUA codons to be replaced by AUU or AUC and make it easier for the CD mechanism to occur, even if the total number 
of Ile codons is quite large. For the runs in Table 1(d), we suppose that the rate of mutation from base $i$ to base $j$ is given 
by $r_{ij} = \alpha \pi_{j}$, where $\pi_j$ is the equilibrium frequency of base $j$. The average mutation rate is  
$u = \sum_{i}\sum_{i\neq j}\pi_{i}r_{ij}$. In the following runs, we varied the frequency $\pi_G$ with the assumption that 
$\pi_C = \pi_G$ and that $\pi_A = \pi_U = 0.5-\pi_G$. We chose $\alpha$ so that $u$ was kept fixed at 0.0001. In these runs, 
\#AUA was 8. Under these conditions the predominant mechanism is AI when the base frequencies are equal. As the mutational bias 
towards GC is increased, the mechanism switches towards predominantly CD, which confirms the important influence of directional 
mutation pressure on the CD mechanism.

  In Table 1(e) we consider variation in $s_{unas}$ whilst $s$ is fixed at 0.02. When $s_{unas}$ is also 0.02, CD is the 
predominant mechanism and UC is rare because of the asymmetry in selection against unassigned and ambiguous codons discussed above. 
The table shows that when $s_{unas}$ is reduced, the UC mechanism becomes more frequent than either AI or CD. In this case there is an 
asymmetry that favours unassigned codons over ambiguous ones. A low value of $s_{unas}$ corresponds to a case where another tRNA for 
the same amino acid would be able to translate the codon after loss of the specific tRNA \cite{yokobori}. In simulations where 
$s_{unas}$ is made even smaller than the values shown, the intermediate unassigned state can be sustained for a very long time. 
This corresponds to a case of reduction of the necessary number of tRNAs for a codon family without change in the genetic code 
(e.g. the number of tRNAs have been reduced to one in all the 4-codon families in mitochondrial genomes).

   The deletion of a tRNA can confer a selective advantage to an individual since it leads to a reduction in the genome size - the 
genome streamlining hypothesis \cite{kurland}. Although Knight et al. \cite{landweber} have argued against this hypothesis, 
it could be an important effect in small genomes like mitochondria, where the length of a tRNA is not negligible 
compared to the total genome length. In the context of our model, selection for reduction in genome length provides an 
advantage to the loss event that can counteract the negative selection from unassigned codons. To investigate this 
effect, we consider the rate of genome replication to be proportional to its length, and that deletion of a tRNA will confer 
a selective advantage dependent on $\delta$, the relative length of a tRNA gene to the total genome length. A genome after the 
loss is a factor $(1-\delta)$ shorter, and its fitness is a factor $1/(1-\delta)$ larger than a genome in which the loss has 
not occurred. Reclinomonas americana is a protist \cite{gray} with a relatively large mitochondrial genome ($L=69034$). Both 
the UGA and the AUA codon retain their canonical assignments in this organism, and the genome still possesses the tRNA-Ile 
specific to the AUA codon. This genome therefore gives us a good idea about the state of the mitochondrial genome before these 
reassignments occurred. The mean length of a tRNA in Reclinomonas is 72, and $\delta = 0.001$. A second example is the human 
mitochondrial genome, which is typical of most animal mitochondrial genomes. This has a length of 16571 and a mean tRNA length 
of 69; hence, $\delta = 0.004$. Table 1(f) shows that as $\delta$ increases the UC mechanism becomes predominant. For the values 
of the selection coefficients used here, there is no significant difference in the frequencies between the runs with 
$\delta=0$ and $\delta=0.001$, but the number of UC cases is significantly higher than either of these when $\delta= 0.004$. 
Unfortunately we do not know the values of the selection coefficients $s$, $s_{unas}$ and $s_{stop}$ in the real case, and 
they may be much smaller than the values we are using in these simulations. This means that this effect could well be significant in 
genomes of the size of Reclinomonas. Some of the important changes that have occurred in the mitochondrial translation 
system appear to have been initiated by tRNA loss. We also note that deletion of the tRNA potentially confers a selective 
advantage simply from the fact that the tRNA no longer needs to be synthesized. This is a contribution to $\delta$ that is 
not proportional to the genome length, but which would produce the same effect as that considered above. 

\subsection*{Discussion and Conclusions}

  These simulations illustrate qualitative trends in the frequencies of the mechanisms. We would like to be able to 
identify the mechanism responsible for each observed reassignment event. We are hindered here by a lack of information 
on the fitness costs associated with the various possible intermediate stages of the codon reassignment process. If 
there is an ambiguous codon, what are the relative affinities of the two tRNAs? If a stop codon is ambiguous, what are 
the relative frequencies of termination and read-through? If a codon is unassigned, is there another tRNA that can step 
in and fulfill the role of the original tRNA? Does the delay to the translation process caused by unassigned codons have 
a significant effect on the fitness, or is it merely the incorrect translation of a codon that would be important?
Even though the fitness costs of an ambiguous or unassigned codon are not yet available, much work has been done to shed 
light on the detailed molecular processes responsible for efficient translation.  

  Recent works have shown the presence of various error correcting mechanisms working at every stage to ensure the 
accuracy of translation \cite{ibba}. When two competing tRNA's with distinct anticodons are available to decode a 
particular codon, as in the ambiguous intermediate case, which of the two eventually wins over would be determined 
by factors such as the efficiency of binding between the charged tRNA and the elongation factor protein \cite{lariv}, 
efficiency of the codon-anticodon pairing at the ribosome and its effect on proper insertion of an amino acid into 
the growing polypeptide chain \cite{ogle}. For the case of an unassigned codon, these factors will determine whether 
another non-cognate tRNA, with an anticodon which mispairs with the corresponding unassigned codon at the wobble position, 
can be recruited and thereby used to extend the polypeptide chain in the absence of a better alternative?  Such work 
was primarily motivated by experiments which indicate that the frequency of mistranslation events is low; 
approx. 1 in 10,000 \cite{kurl}. Mistranslation, which can arise either due to the mispairing of codon and anticodon 
or due to mischarging of a tRNA, is qualitatively different from ambiguous translation of a codon considered here. 
During ambiguous translation, both the tRNAs in question have anticodons that can 'correctly' pair with the same codon, 
and there is no reason to say that one or the other is a mistranslation. Also, the ambiguous state seems to arise through 
the gain of function of a new tRNA (usually by an anticodon change) not through the breakdown of the accuracy of translation 
of the original tRNA. Another relevant point related to mistranslation is that it causes a loss of fitness since it results 
in non-optimal proteins. The severity of the load imposed on an organism by mistranslation will depend on which types of 
amino acid substitutions result most often from mistranslations. The structure of the canonical genetic code appears to 
minimize the load due to mistranslation by placing similar amino acids on neighboring codons in the table \cite{freeland1,freeland2}. 
Synonymous codons might also be subject to different frequencies of mistranslation. This would mean that 
replacement of all occurrences of a codon by another synonymous codon (as happens in the CD mechanism) might not be 
strictly neutral. This was not considered in our model, since we assumed that any fitness effects due to synonymous 
changes were negligible compared to non-synonymous ones. 

  We would like to be able to predict which mechanisms of codon reassignment occur most frequently in real genomes. 
Unfortunately, real parameter values lie outside the range that can be simulated with the model: genomes should be 
much longer, population sizes larger, mutation rates lower, and selection coefficients most likely should be smaller.
For simplicity, our model also assumes that all non-synonymous changes have equal fitness effect. Allowing a distribution 
of fitness effects across sites would influence the relative frequencies of the mechanisms to some extent. Despite these 
uncertainties, we can make a few general statements. 
	
    Firstly, all these mechanisms become more difficult when genome sizes are larger. This is presumably the main reason for the 
larger number of observed changes in mitochondria (with small genomes and high mutation rates) than in nuclear genomes. The CD 
mechanism is most likely in cases where there are initially few copies of the codon to be reassigned. Thus, it could well apply for 
stop codon reassignments, and for cases when there is a strong mutational bias that causes the reduction in number of one of a 
family of synonymous codons. It seems less likely than the other mechanisms for most reassignments other than stop codons. The fact 
that so many real cases of reassignments are now known suggests to us that our intuition regarding the severity of ambiguous and 
unassigned codons may have been wrong. Our results suggest that many of the cases where changes have occurred may have passed 
through one of these two intermediates. 

  There is another line of argument that favours the AI and UC mechanisms over the CD. The gain and loss events that lead to a 
reassignment are independent, rare events involving distinct tRNAs or RFs. According to the CD mechanism, the two events occur by 
chance during a period in which the codon disappears. This means that if a gain (or loss) occurs that affects a particular codon, 
then the corresponding loss (or gain) must occur for the same codon soon afterwards, and both these changes must spread through the 
population as neutral mutations. There is no particular reason why the two rare events should be associated with the same codon. The 
AI and UC mechanisms do not suffer from this problem. In both cases, after the first gain (or loss) there is a selective advantage 
for the loss (or gain) for the same codon. Thus the gain and loss events are linked by selection in these two mechanisms, but not 
in the CD mechanism. In the simulations, both gain and loss occurred at a rate of 0.1 per population per generation. This high 
rate was used to ensure that a supply of gain and loss mutations was available, some of which were able to go to fixation in the 
population within a reasonable simulation time.  Consequently, it is easy for both gain and loss events to be present in the 
population at the same time and to be fixed neutrally. This favors the CD mechanism. If the gain and loss rate were lower, 
we would expect a reduction in the frequency of CD relative to AI and UC.

    It is worth noting that we set the rate of the gain and loss events to be equal. If these rates were not equal, this would 
affect the relative frequencies of AI and UC mechanisms in an obvious way. A loss event seems simpler than a gain in most cases. 
It is not difficult to delete a gene, whereas most of the gains discussed here involve creation of a modified base. This 
means that the enzyme for base modification has to appear from somewhere and learn to recognize the appropriate tRNA. For 
these reasons we might expect the loss rate to be higher than the gain rate in real life, and hence the UC mechanism would 
be more frequent. 
  
  The CC mechanism was never common for any of the parameter sets used. However, we would expect this to become more frequent 
in longer genomes, where CD is unlikely and where there would be large penalties for AI and UC states.

   Directional mutation pressure is beneficial for CD, UC and AI mechanisms. In the CD mechanism,while not essential for ensuring 
codon disappearance, it makes this more likely in longer genomes containing a large number of such codons. Also, by considerably 
reducing the number of codons undergoing reassignment via the AI or UC mechanisms, it helps offset the deleterious effect of 
ambiguous translation or an unassigned codon. 
	
	In summary, this paper makes several important contributions to the understanding of the codon reassignment process. By 
introducing the gain-loss model of reassignment, we have provided a unified framework for discussing two of the most frequently 
cited mechanisms (CD and AI) of code change. Expressing the problem in this way highlights the similarity with the theory of 
compensatory mutations that has been studied in other contexts. Within our framework, the UC and CC mechanisms appear as natural 
alternatives to CD and AI. The framework is sufficiently general to apply to many different reassignment events, despite the fact 
that the molecular events causing the gains and losses are different in every case. In addition, we have presented a well-defined 
population genetics model, shown that all four mechanisms occur in the simulations, and discussed the influence of the parameters 
on the relative frequency of the mechanisms. We are now investigating the many known cases of codon reassignment to determine 
which mechanisms appear to have been most frequent in real organisms.

\subsection*{Acknowledgements}
 
	This work was supported by the Canada Research Chairs.



\begin{thebibliography}{99}

\vspace{-0.1in}

\bibitem{knight} Knight, R.J., Freeland, S.J., \& Landweber, L.F. (2001) {\it Nature Reviews Genetics} 
{\bf 2}, 49-58.

\bibitem{osawa} Osawa, S. \& Jukes, T.H. (1989) {\it J. Mol. Evol.} {\bf 28}, 271-278 (1989).

\bibitem{yarus1} Schultz, D.W., and Yarus, M. (1994) {\it J. Mol. Evol.} {\bf 235}, 1377-1380;
Schultz, D.W., and Yarus, M. (1996) {\it J. Mol. Evol.} {\bf 42}, 597-601.

\bibitem{matsugi} Lovett, P.S., Amboulos, N.P.J., Mulbry, W., Noguchi, N., \& Rogers, E.J.,
(1991) {\it J. Bacteriol.} {\bf 173}, 1810-1812; Matsugi, J., Murao, K., \& Ishikura, H.,
(1998) {\it J. Biochem.} {\bf 123}, 853-858.

\bibitem{santos} Santos, M.A.S., Ueda, T., Watanabe, K. \& Tuite, M. (1997) {\it Mol. Microbiol},
{\bf 26}, 423-431 and references therein.

\bibitem{muramatsu} T. Muramatsu, S. Yokoyama, N. Horie, A. Matsuda, T. Ueda, Z. Yamaizumi,
Y. Kuchino, S. Nishimura \& Miyazawa, T. (1988) {\it J. Biol. Chem.} {\bf 263}, 9261-9267.

\bibitem{tang} Tang, B., Boisvert, P. \& Higgs, P.G. (2004) Comparison of tRNA and rRNA 
phylogenies in proteobacteria. ({\it submitted})

\bibitem{gray} Gray, M.W., Lang, B.F., Cedergren, R., Golding, G.B., Lemieux, C., {\it et.al.}
(1998) {\it Nucleic Acids Res.} {\bf 26}, 865-878.

\bibitem{tomita} Tomita, K., Ueda, T., Ishiwa, S., Crain, P.F., McCloskey, J.A., \& Watanabe, K.
(1999) {\it Nucleic Acids Res.} {\bf 27}, 4291-4297.

\bibitem{yokobori} Yokobori, S. Suzuki, T \& Watanabe, K. (2001) {\it J. Mol. Evol.} {\bf 53},
314-326.

\bibitem{matsuyama} Matsuyama, S., Ueda, T., Crain, P.F., McCloskey, J.A., \& Watanabe, K.
(1998) {\it J. Biol. Chem.} {\bf 273}, 3363-3368; 

\bibitem{kondow} Yokobori, S., Ueda, T., \& Watanabe, K. (1993) J. Mol. Evol. {\bf 36}, 1-8;
Kondow, A., Suzuki, T., Yokobori, S., Ueda, T., \& Watanabe, K. (1999) {\it Nucleic Acid Res.}
{\bf 27}, 2554-2559.

\bibitem{kimura} Kimura, M. (1985) {\it J. Genet.} {\bf 64}, 7-19.

\bibitem{higgs} Higgs, P.G. (1998) {\it Genetica} {\bf 102/103}, 91-101.  

\bibitem{savill} Savill, N.J., Hoyle, D.C. \& Higgs, P.G. (2001) {\it Genetics} {\bf 157}, 399-411. 

\bibitem{wong} Wong, J.T.F. (1975) {\it  Proc. Natl. Acad. Sci. USA} {\bf 72}, 1909-1912; Wong, J.T.F. (1976) {\it Proc. Natl. Acad. Sci. USA} {\bf 73}, 2336-2239.

\bibitem{szathmary} Szathmary, E. (1991) {\it J. Mol. Evol.} {\bf 32}, 178-182.

\bibitem{freeland1} Freeland, S.J., \& Hurst, L.D. (1998) {\it J. Mol. Evol.} {\bf 47} 238-248.

\bibitem{freeland2} Freeland, S.J., Knight, R.D., Landweber, L.F., \& Hurst, L.D. (2000) {\it Mol. Biol. Evol} 
{\bf 17} 511-518.

\bibitem{ardell} Ardell, D.H. \& Sella, G (2001) {\it J. Mol. Evol.} {\bf 53}, 269-281; Sella, G., \& Ardell, D.H. (2002) {\it J. Mol. Evol.} {\bf 54}, 638-651.

\bibitem{kurland} S.G. Andersson and C.G. Kurland, (1991) Mol. Biol. Evol. {\bf 8}, 530-544;
{\it ibid}, (1995) Biochem. Cell. Biol. {\bf 73}, 775-787.

\bibitem{landweber} Knight, R.D., Landweber, L.F., \& Yarus, M. (2001) {\it J. Mol. Evol.} {\bf 53}, 
299-313.

\bibitem{ibba} Ibba, M. \& Soll, D. (1999) {\it Science} {\bf 286}, 1893-1897.

\bibitem{lariv} Lariviere, F.J, Wolfson, A.D., \& Uhlenbeck, O.C. (2001) {\it Science} {\bf 294}, 165-168. 

\bibitem{ogle} Ogle, J.M., Carter, A.P. \& Ramakrishnan, V. (2003) {\it Trends. Biochem. Sci.} {\bf 28}, 259-266.  

\bibitem{kurl} Kurland, C.G. (1992) {\it Annu. Rev. Genet.} {\bf 26}, 29-50. 

\end{thebibliography}
\end{document}